\documentclass[conference]{IEEEtran}
\usepackage{cite}
\usepackage[pdftex]{graphicx}
\graphicspath{{../pdf/}{../jpeg/}}
\DeclareGraphicsExtensions{.pdf,.jpeg,.png}
\usepackage{amsmath,amsfonts,amssymb}
\usepackage{array}
\usepackage{mdwtab}
\usepackage[font=footnotesize]{subfig}
\usepackage{hyperref}
\usepackage{color}
\usepackage{url}
\usepackage{booktabs}
\usepackage[ruled,vlined,linesnumbered]{algorithm2e}
\usepackage{threeparttable}
\usepackage{footnote}
 
\SetKwProg{Fn}{Function}{}{}
\usepackage{caption}
\usepackage{paralist}
\hypersetup{
    colorlinks   = true,
}

\begin{document}
\title{ BLASX: A High Performance Level-3 BLAS Library for Heterogeneous Multi-GPU Computing \vspace{-0.35in}}
\author{
\IEEEauthorblockN{
\href{mailto:linnan.wang@gatech.edu}{Linnan Wang}\IEEEauthorrefmark{1},
\href{mailto:wwu12@vols.utk.edu}{Wei Wu}\IEEEauthorrefmark{2},
\href{mailto:xj@princeton.edu}{Jianxiong Xiao}\IEEEauthorrefmark{3}, and
\href{mailto:yyang@nec-labs.com}{Yi Yang}\IEEEauthorrefmark{4}
}
\IEEEauthorblockA{
    \IEEEauthorrefmark{1}
    Georgia Institute of Technology
}
\IEEEauthorblockA{
    \IEEEauthorrefmark{2}
    The University of Tennessee, Knoxville
}
\IEEEauthorblockA{
    \IEEEauthorrefmark{3}
    Princeton University
    }
\IEEEauthorblockA{
    \IEEEauthorrefmark{4}
    NEC Laboratory
    }
}
\maketitle
\begin{abstract}
Basic Linear Algebra Subprograms (BLAS) are a set of low level linear algebra kernels widely 
adopted by applications involved with the deep learning and scientific computing.
The massive and economic computing power brought forth by the emerging GPU architectures drives
interest in implementation of compute-intensive level 3 BLAS on multi-GPU systems.
In this paper, we investigate existing multi-GPU level 3 BLAS and present that 
1) issues, such as the improper load balancing, inefficient communication, insufficient GPU 
stream level concurrency and data caching, impede current 
implementations from fully harnessing heterogeneous computing resources;
2) and the inter-GPU Peer-to-Peer(P2P) communication remains unexplored.
We then present BLASX: a highly optimized multi-GPU level-3 BLAS. 
We adopt the concepts of algorithms-by-tiles treating a matrix tile as the basic data unit 
and operations on tiles as the basic task. Tasks are guided with a dynamic asynchronous 
runtime, which is cache and locality aware. The communication cost under BLASX becomes 
trivial as it perfectly overlaps communication and computation across multiple streams 
during asynchronous task progression. It also takes the current tile 
cache scheme one step further by proposing an innovative 2-level hierarchical tile cache, 
taking advantage of inter-GPU P2P communication. As a result, linear speedup is observable with 
BLASX under multi-GPU configurations; and the extensive benchmarks demonstrate that BLASX consistently 
outperforms the related leading industrial and academic projects such as cuBLAS-XT, SuperMatrix, 
MAGMA and PaRSEC.
\end{abstract}


\footnotetext{BLASX is publicly available at https://github.com/linnanwang/BLASX }
\vspace{-0.1in}
\section{Introduction}
Matrix scaling, additions and multiplications are basic operations in linear algebra libraries, 
scientific simulations and deep learning. Offloading these operations to the BLAS library, 
which is the case in general practices, can substantially improve the application performance 
due to the architecture specific optimizations inside BLAS kernels. This leads BLAS to become the 
standard building blocks for performing low level matrix operations in applications. Hence, 
the BLAS library directly affects the application performance. In the last few years, 
the evolving GPU architectures, NVIDIA's Kepler and Maxwell in particular, feature thousands
of stream processors, proven to be extremely efficient in computing level 3 BLAS.

While a multi-GPU system offers appealing high performance, using it entails nontrivial effort. 
A multi-GPU system typically consists of at least one CPUs connected with peripheral GPUs on 
the PCI-E. GPU operates on its private onboard RAM while CPU operates on the host RAM. 
For GPUs sharing the same I/O hub, they can directly communicate via PCI-E switch referred to
as GPU P2P access. The following factors need to be considered to fully utilize such architecture: 
(1) nowadays GPUs of different architectures represent divergent computing 
capabilities; and even the realtime performance of a GPU varies with respect to factors 
such as kernel saturation and GPU occupancy, all of which pose a great challenges to load 
balancing. (2) Minimizing and overlapping communication is key to achieve
high performance. (3) Reducing the CPU-GPU communication to the GPU-GPU communication
further improves the communication and energy efficiency. Unfortunately, 
our study indicates the existing multi-GPU level-3 BLAS fail to optimize toward these factors,
thereby delivering sub-optimal performance.

In this paper, we present the BLASX: a high-performance level-3 BLAS library for heterogeneous 
multi-GPU systems. We address the load balancing with a dynamic scheduling runtime, which 
handles task level workload variations, single GPU realtime performance variations and speed 
discrepancies among heterogeneous multi-GPUs. BLASX adopts a novel two level hierarchical 
tile caches to explore the tile temporal locality, in which we consider the GPU onboard RAM 
as the L1 tile cache and the combined multi-GPU RAMs as the L2 tile cache. The L1 tile
cache minimizes global communication; and the L2 tile cache successfully reduces 
the CPU-GPU communication to the GPU-GPU communication. In implementing this hierarchical tile 
caches, we propose a new LRU algorithm to accommodate the asynchronous task progression and 
a new cache coherence protocol to ensure the data consistency on multi-GPUs. BLASX also
optimizes the communication/computation overlapping on GPU streams so that the communication
cost is negligible. Finally, BLASX offers backward compatibility to the vast existing CPU BLAS based 
applications; thereby all the details, such as workload balancing, data caching, communication 
overlapping and memory management, can be ignored by library users.

We evaluate BLASX on two multi-GPU systems, Everest and Makalu (TABLE \ref{Machines}), against 
the related leading academic and industrial projects including cuBLAS-XT, SuperMatrix, MAGMA and PaRSEC.
BLASX consistently outperforms the academic implementations on Everest with 3 NVIDIA Kepler 
K40c. In contrast to the highly optimized NVIDIA cuBLAS-XT, BLASX demonstrates on average 
25\% performance gain and 200\% less communication volume. Makalu features 2 Kelper K40 and 2 Maxwell 
TITAN X. BLASX successfully tackles the heterogeneity and demonstrates linear speedup; 
whereas other libraries such as cuBLAS-XT, MAGMA and SuperMatrix suffers from poor scalability.

We organize the remaining paper as follows.
Section II analyzes the background and related works; and section III briefly reviews 
the L3 BLAS tile algorithms. In Section IV, we elaborate the detailed design and implementations 
of BLASX including two level hierarchical tile caches and the scheduling runtime. 
We also present solutions to specific questions such as amortizing high frequency memory allocation 
and deallocation, communication and computation overlapping. The comprehensive evaluations 
of BLASX against existing state-of-art implementations are presented in the Section V. 
Finally, we conclude at Section VI.

\vspace{-0.12in}
\section{Background and Related Work}
\begin{figure}
\vspace{-0.10in}
\centering
\mbox{
\subfloat[SuperMatrix]{\includegraphics[width=2.5in]{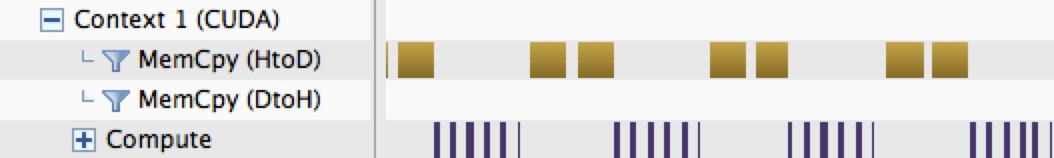}
\label{SuperMatrix}}
}
\hfil
\mbox{
\subfloat[StarPU]{\includegraphics[width=2.5in]{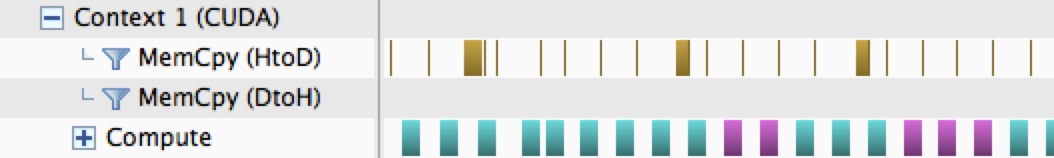}
\label{StarPU}}
}
\hfil
\mbox{
\subfloat[cuBLAS-XT]{\includegraphics[width=2.5in]{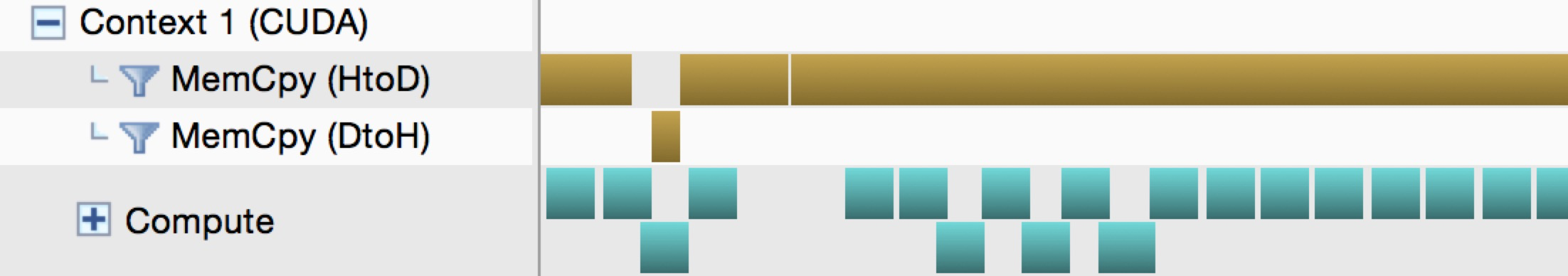}
\label{cuBLAS-XT}}
}
\hfil
\mbox{
\subfloat[BLASX]{\includegraphics[width=2.5in]{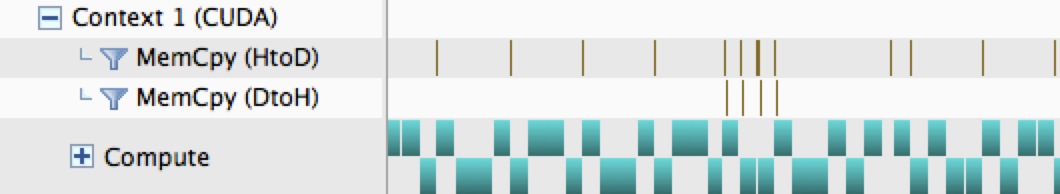}
\label{BLASX}}
}
\caption{A snapshot of single GPU DGEMM execution profile from SuperMatrix, StarPU, cuBLAS-XT and BLASX.}
\label{GPU_profile}
\vspace{-0.25in}
\end{figure}

There are three levels of BLAS, divided with respect to the complexity of operations. 
Level-1 (L1) BLAS targets vector operations in $O(n)$ such as vector dot products and 
vector norms. Level-2 (L2) BLAS targets matrix-vector operations in $O(n^{2})$ such as
matrix-vector multiplication. Level-3 (L3) BLAS \cite{Dongarra} targets matrix operations 
in $O(n^{3})$ time such as matrix-matrix multiplications. The focus of this research is 
on L3 BLAS, which uses General Matrix Multiplication (GEMM) as the primary building 
block for the routines within the category. Therefore, the task of improving the performance 
of L3 BLAS can be reduced to the GEMM speed.

The massive but economic TFLOPS brought forth by the evolving GPU architectures drives 
interests in the various implementations of multi-GPU L3 BLAS. SuperMatrix \cite{SuperMatrix} 
is one of the pioneers of matrix operation parallelization on SMP multicores, however it 
provides limited support on GPUs. The key insight of SuperMatrix is that a matrix can be 
factorized into a set of tiles. The Tomasolu algorithm \cite{Tomasulo} subsequently 
schedules these tiles in the out-of-order fashion. Fig.\ref{SuperMatrix} demonstrates that
SuperMatrix suffers from costly nonoverlapped CPU-GPU data transfers. StarPU 
provides a centralized interface to various accelerator technologies \cite{StarPU}. 
In contrast to Supermatrix, StarPU supports versatile scheduling algorithms such as work 
stealing \cite{work_stealing} and priority scheduling \cite{priority_scheduling} while requiring 
manual annotations to optimize under a specific problem. The insufficient communication/computation overlapping and the 
low GPU saturation in Fig.\ref{StarPU} demonstrate the suboptimal DGEMM implementation
in StarPU. MAGMA \cite{Nath} is another multi-GPU linear algebra library with incomplete 
LAPACK and BLAS support. It is a heavily hand tuned library relying on a static load balancer,
which degrades MAGMA's performance on heterogeneous multi-GPU systems. 
Direct Acyclic Graph (DAG) scheduling has seen a revival in the recent years as it can naturally integrate
with tile algorithms. PaRSEC \cite{PaRSEC} is a leading DAG scheduling runtime for 
dense linear algebraic operations. Building DAGs at runtime and scheduling tasks within DAGs, however, 
can be a huge cost for the small scale L3 BLAS operations. PaRSEC also assumes constant 
workload on each fine-grained task and constant speed on each GPU. It is possible 
to have workload variations and the GPU kernel saturation also affects the actual execution 
speed. In addition, PaRSEC only exploits tile reusing within a single GPU; Caching on 
multiGPU memory spaces by taking advantages of GPU-GPU P2P communication still remains unexplored.
None of the aforementioned libraries are backward compatible to legacy CPU BLAS. As an reaction 
to the market, NVIDIA released a commercial multiGPU L3 BLAS, cuBLAS-XT \cite{cuBLASXT}, declaring it to be 
backward compatible when using the NVBLAS wrapper. cuBLAS-XT consistently 
moves tiles on demand into GPU RAM so that it can compute a large scale problem 
with a few MB of GPU RAM. Although major communication is overlapped, it does 
not address tile caching; and this aggressive on demand communication pattern extremely overloads 
the PCI-E as shown by the contiguous yellow blocks in Fig.\ref{cuBLAS-XT}. 

In summary, these libraries cannot deliver the optimal performance due to following issues:
1) insufficient communication/computation overlapping subjects SuperMatrix and StarPU
to suboptimal performance. 2) excessive communication in cuBLAS-XT overloads the PCI-E
dragging down the overall performance. 3) low GPU occupancies in SuperMatrix and StarPU
indicate partial GPU utilization.
4) efficient GPU-GPU P2P communication remains unexplored. We observe that the 
average throughput of CPU-GPU communication is 6.54 GB/S while the GPU-GPU is 7.80 GB/S. 
5) static scheduling in the cuBLAS-XT and MAGMA cannot tackle the hardware heterogeneity.
BLASX successfully resolves these issues as 
demonstrated in Fig.\ref{BLASX}; please refer to \textit{Performance Evaluation} for detailed 
discussion.

\vspace{-0.1in}
\section{A Review of L3 BLAS Tile algorithms}
In this section, we give an overview of L3 BLAS tile algorithms. L3 BLAS is intended for $O(n^3)$ matrix 
operations including General Matrix Multiplication (GEMM), symmetric rank-k update (SYRK), 
symmetric rank-2k update (SYR2K), triangular matrix multiplication (TRMM), symmetric matrix 
multiply (SYMM), and triangular solve with multiple right hand side (TRSM). Since the Hermitian 
matrix multiplication (HEMM), Hermitian rank-k update (HERK), and Hermitian rank-2k update 
(HER2K) are the complex counterparts of GEMM, SYRK and SYR2K respectively, we omit their 
discussion in this paper.

\vspace{-0.12in}
\subsection{Representing Matrix as Tiles}
\vspace{-0.05in}
The tile algorithm logically partitions a matrix into its tiled representation. 
Given tile size $T$ in a matrix of size $ N \times M $, it creates $\lfloor N/T \rfloor \times \lfloor M/T \rfloor $ 
square tiles of size $ T \times T $ and $(\lceil N/T \rceil \times \lceil M/T \rceil ) - (\lfloor N/T \rfloor \times \lfloor M/T \rfloor)$ non-square tiles.
Furthermore, the algorithm treats tiles, uniquely indexed by row and column, as the basic 
elements in a matrix in lieu of scalars. Operations on matrices are subsequently reduced 
to operations on tiles. As our focus is the L3 BLAS, we assume the tile indices 
of the output matrix are $[i, j]$, the tile indices of the matrix to the left side of 
multiply operator are $[i, k]$ and the tile indices of matrix to the right of multiply 
operator are $[k, j]$. Hence tile indices $i$ and $j$ uniquely identify a tile, $\mathbf{C_{ij}}$, 
in the output matrix while the upper bound of $k$ represents the computational intensity to solve the $\mathbf{C_{ij}}$.

\begin{table}[!t]
\caption{GEMM percentages at 3 different matrix sizes N.}
\centering
\label{GEMM_Percentages}
\begin{tabular}{c c c c}
    \toprule
    \textbf{Routines} & \textbf{N=5K} & \textbf{N=10K} & \textbf{N=20K} \\ \midrule
      SYRK            &      74.5\%         &       86.3\%        &   92.8\%  \\
      TRSM            &      68.5\%         &       80.4\%        &    89\%   \\
      TRMM            &       69\%          &       81.5\%        &   92.8\%  \\
      SYR2K           &      74.4\%         &       85.4\%        &   92.9\%  \\
      SYMM            &      71.7\%         &       84.9\%        &   92.1\%  \\ \bottomrule
\end{tabular}
\vspace{-0.15in}
\end{table}

\vspace{-0.12in}
\subsection{Traditional GEMM based L3 BLAS Tile Algorithms}
\vspace{-0.05in}
The traditional tile implementation of L3 BLAS on the CPU relies on a highly optimized GEMM and a small amount 
of L1 and L2 BLAS \cite{Goto} or other L3 BLAS routines. The following equations illustrate the none-transpose, upper 
triangular cases of L3 BLAS tile algorithms: GEMM, SYRK, TRSM, TRMM, SYR2K and SYMM, respectively:

\begin{subequations}
\vspace{-0.27in}
 \label{L3_Tile_BLAS}
 \begin{align}
  \mathbf{C_{ij}} &= \alpha \sum \limits_{k=0}^{z} \mathbf{A_{ik}} \mathbf{B_{kj}}+\beta \mathbf{C_{ij}} \label{GEMM_e}               \\
  \mathbf{C_{ij}} &= \alpha \sum \limits_{k=0}^{z} \mathbf{A_{ik}} \mathbf{A_{jk}^\intercal}+\beta \mathbf{C_{ij}} \label{SYRK_e}     \\
  \mathbf{C_{ij}} &= \alpha \mathbf{A_{ii}^{-1}} \left( \mathbf{B_{ij}} - \sum \limits_{k=i+1}^{z} \mathbf{A_{ik}} \mathbf{C_{kj}} \right) \label{TRSM_e} \\    
  \mathbf{C_{ij}} &=\alpha \sum \limits_{k=i+1}^{z} \mathbf{A_{ik}} \mathbf{C_{kj}} + \mathbf{A_{ii}C_{ij}} \label{TRMM_e}              \\
  \mathbf{C_{ij}} &=\alpha \sum \limits_{k=0}^{z} \mathbf{A_{ik}} \mathbf{B_{jk}^\intercal} + \alpha \sum \limits_{k=0}^{z} \mathbf{B_{ik}} \mathbf{A_{jk}^\intercal} +\beta \mathbf{C_{ij}} \label{SYR2K_e}\\
  \mathbf{C_{ij}} &= \alpha \sum \limits_{k=0}^{i} \mathbf{A_{ki}^\intercal} \mathbf{B_{kj}} + \alpha \sum \limits_{k=i+1}^{z} \mathbf{A_{ik}} \mathbf{B_{kj}} + \beta \mathbf{C_{ij}} \label{SYMM_e} 
 \end{align}
\end{subequations}
\vspace{-0.18in}

\vspace{-0.12in}
\subsection{A simple trick to Matrix Transpose}
\vspace{-0.05in}
The transpose of a tiled matrix requires a transpose on each tiles in addition to swapping the 
tiles $\mathbf{A_{ij}}$ and $\mathbf{A_{ji}}$ to be consistent with the definition of matrix transpose. 
The concept is as follows:\\
\vspace{-0.10in}
\begin{equation*}
\begin{aligned}
&
\begin{bmatrix}
\mathbf{A_{00}} & \mathbf{A_{01}} \\
\mathbf{A_{10}} & \mathbf{A_{11}}
\end{bmatrix}
\begin{bmatrix}
\mathbf{B_{00}} & \mathbf{B_{01}} \\
\mathbf{B_{10}} & \mathbf{B_{11}}
\end{bmatrix}^\intercal 
=
\begin{bmatrix}
\mathbf{A_{00}} & \mathbf{A_{01}} \\
\mathbf{A_{10}} & \mathbf{A_{11}}
\end{bmatrix}
\begin{bmatrix}
\mathbf{B_{00}^\intercal} & \mathbf{B_{10}^\intercal} \\
\mathbf{B_{01}^\intercal} & \mathbf{B_{11}^\intercal}
\end{bmatrix}
\end{aligned}
\end{equation*}
This simple trick significantly facilitates the matrix transpose. Rather than physically 
transposing the entire matrix, we can retrieve the tile $\mathbf{A_{ji}}$ and transpose 
the tile inside the BLAS kernel to implicitly transpose the matrix.

\vspace{-0.12in}
\subsection{The GEMM Dominant L3 BLAS}
\vspace{-0.05in}
TABLE \ref{GEMM_Percentages} presents the GEMM percentages in L3 BLAS with respect to 3 square
matrix sizes. The percentages of GEMM, as indicated in \ref{GEMM_e} to \ref{SYMM_e}, increase
with the size of matrices to a point that the entire computation eventually be dominated by 
the GEMM kernel, achieving GEMM-like performance.

\section{BLASX: A Multi-GPU L3 BLAS Library with a Locality Aware Dynamic Scheduling Runtime}

We organize this section as follows. We begin by introducing a set of new L3 BLAS tile 
algorithms for heterogeneous multi-GPUs. Subsection A elaborates the L3 BLAS task-ization;
Subsection B elaborates a novel two-level hierarchical tile cache on multi-GPU RAMs; 
Subsection C elaborates our locality aware scheduling runtime covering load balancing,
scheduling infrastructure, and scheduling strategies; 
Subsection D elaborates communication/computation overlapping and GPU out-of-core operations. 
In the end, we present a novel heap design to amortize the overhead introduced by 
high-frequency GPU memory allocation/deallocation.

\vspace{-0.1in}
\begin{algorithm}
\caption{A new L3 BLAS tile algorithm for heterogeneous multi-GPUs}
\label{BLASX_alg}
\DontPrintSemicolon
\KwData{$\mathbf{A}, \mathbf{B}$ and $\mathbf{C}$}
\KwResult{$\mathbf{C}$}
\Begin{
$None Blocking Task Queue (TQ) \leftarrow init(\mathbf{A}, \mathbf{B}, \mathbf{C})$\\
$Cache Coherence Procotol (CCP) \leftarrow init()$\\
	\For{$ gpu \in  GPUs$} {
	    $SpawnThread(ComputingKernel, [TQ, CCP])$\\
	}
	$AllThreadsJoin()$
}
\hfill \\
\Fn{$ComputingKernel$ ($TQ, CCP$)}{
    $BindToCore(self)$ \\
    \While{$TQ \neq \emptyset$ }{ \
        \For{$ RS \in  ReservationStation (RS)$}{
            \If{$RS = \emptyset$} {
                $task \leftarrow {Dequeue(TQ) \mbox{ or } WorkStealing()}$ \\
                $priortiy \leftarrow CalculatePriority(task)$ \\
                $RS \leftarrow \{task, priortiy\} $\\
            }
        }
        $StreamsSynch()$\\
        $ReaderUpdate()$\\
        \For{$ k $}{
            \For{$ task \in$ Top 4 Prioritized Tasks in RS}{
                $i \leftarrow task$, $j \leftarrow task$\\
                $stream\_idx \leftarrow task$\\
                $\mathbf{A_{ik}} \leftarrow CCP(\&(task\rightarrow\mathbf{A_{ik}}))$\\
                $\mathbf{B_{kj}} \leftarrow CCP(\&(task\rightarrow\mathbf{B_{kj}}))$\\
                $SetStream(stream\_idx)$\\
                $AsyncBLAS(\mathbf{A_{ik}}, \mathbf{B_{kj}}, \mathbf{C_{ij}})$\\   
            }
        }
    }
}
\end{algorithm}
\vspace{-0.15in}

Alg.\ref{BLASX_alg} describes the skeleton of the proposed L3 BLAS algorithms for heterogeneous 
multi-GPUs. Lines 1 to 6 indicate the general runtime procedures; and 
lines 8 to 25 indicate the GPU specifics during the computation. In lines 1 to 6, the runtime 
initializes a hierarchical tile cache and a global non-blocking task queue. Then 
a CPU thread is spawned for each GPU to submit instructions. The threads join together 
after the global task queue depletes. In lines 8 to 25, GPUs concurrently retrieve tasks
and interleave them via multi-streams to overlap the communication during the asynchronous progression. 
The line 13 indicates two ways of task retrieval, either from global task queue or working stealing.
A GPU steals tasks from other Reservation Stations (RS), which only happens when the GPU 
exhausts tasks on RS while the global queue is also empty. Lines 22 and 23 reflect tile reuses with 
the proposed tile caches. Line 25 suggests an asynchronous L3 BLAS kernel invocation, 
the type of which is dictated by Eq.\ref{L3_Tile_BLAS} given the tile indices $i$, 
$j$ and $k$. Lines 18, 19 and 24 indicate the communication/computation overlapping.
Although we implement this algorithm on NVIDIA GPUs, we can easily migrate it to work on 
other accelerator technologies such as Intel Xeon Phi and AMD FirePro.

\begin{figure}[!t]
\centering
\includegraphics[width=2.5in]{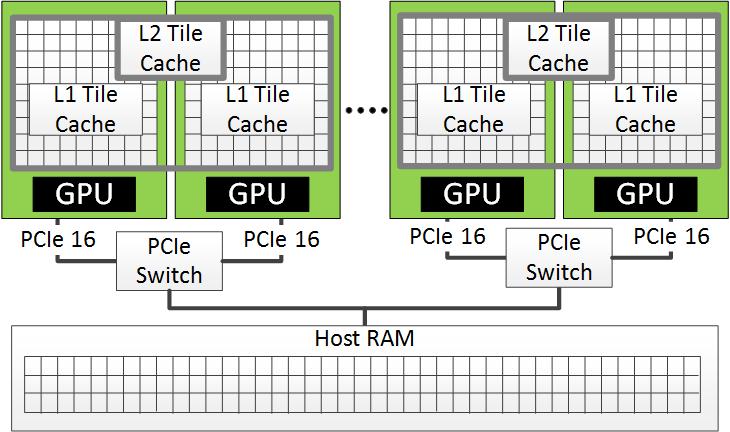}
\caption{The proposed two levels hierarchical tile caches. The L1 tile cache is the GPU onboard RAM; 
the L2 tile cache is the combined RAMs of GPUs on the same PCI-E switch. Each block holds a 
tile on a contiguous independent memory segment.}
\label{tile_caching}
\vspace{-0.25in}
\end{figure}

\vspace{-0.12in}
\subsection{Taskizing L3 BLAS}
\vspace{-0.05in}

We define the task as solving a $\mathbf{C_{ij}}$ in Eq.\ref{BLASX_alg}.
Tile algorithms yield the insight to systematically break down the output matrix $\mathbf{C}$ 
into a set of tiles $\mathbf{C_{ij}}$, the computation of which, involves reading $\mathbf{A_{ik}}$, 
$\mathbf{B_{kj}}$, $\mathbf{C_{ij}}$ and an independent write of $\mathbf{C_{ij}}$. Hence 
concurrently solving the $\mathbf{C_{ij}}$ is data hazards free. Eq.\ref{L3_Tile_BLAS} 
also indicates that the workload of $\mathbf{C_{ij}}$ in non-GEMM routines varies 
according to the upper bound of $k$. In summary, a task has 3 notable properties by following definition:
\begin{itemize}
\vspace{-0.1in}
\item Reading the inputs for a task is data dependency free.
\item Concurrent writing a task's output is data race free.
\item The workload of each task varies.
\end{itemize}
\vspace{-0.1in}
Given a matrix $\mathbf{C}$ of size $M \times N$ and tile size T, the degree of parallelism is as follows:
\vspace{-0.1in}
\begin{equation} 
degree\mbox{-}of\mbox{-}parallelism = \lceil M/T \rceil * \lceil N/T \rceil
\vspace{-0.1in}
\end{equation}
In our implementation, a task holds the necessary metadata to solve a $\mathbf{C_{ij}}$ such as tile
indices $i$, $j$ and $k$, the dimensions of $\mathbf{C_{ij}}$ , and its host address. 
The runtime virtually slices a matrix and stores the tile metadata in tasks. Consequently, taskizing a L3 
BLAS does not require significant additional memory.

\vspace{-0.12in}
\subsection{Two Level Hierarchical Tile Caches}
\vspace{-0.05in}

Fig.\ref{tile_caching} demonstrates the structure of separate memory spaces on the multiGPU system.
Each GPU equips with a private RAM; and CPUs share the host RAM. Nowadays, GPU
RAM can be up to $12$ GB while a single double precision matrix of size  $32768*32768$ is $8.6$ GB. 
The relatively small GPU RAM capacity limits the GPU in-core computing from handling the large scale L3 BLAS. 
One solution using tile algorithms is to dissect a large L3 BLAS into smaller ones, 
and solve them in succession. At each time, we move in tiles from the host on demand. 
Frequent GPU off-chip memory access, however, overloads the PCI-E and degrades the 
performance to be suboptimal. Meanwhile it is possible to reuse certain tiles in separate 
tasks as indicated by Eq.\ref{L3_Tile_BLAS}. Therefore, exploiting the tile 
temporal locality by caching the most frequently used ones on the GPU RAM is necessary.

We present a novel two level fully associative tile caches in Fig.\ref{tile_caching}. 
The L1 tile cache is implemented using GPU onboard RAM; the L2 tile cache is implemented using the combined 
memory spaces of GPUs, which share the same I/O hub. The L1 tile cache hit enables direct tile 
reuse; the L2 tile cache hit reduces the CPU-GPU communication to GPU-GPU communication by 
retrieving the tile from the hardware adjacent GPU. The rational of implementing the L2 cache 
are twofold: 
(1) GPU P2P communication better saturates PCI-E delivering at least 6 GB/s performance. \cite{P2P} 
(2) The comparably faster GPU RAM reduces the latency of data fetching. 
Therefore it is more cost effective to retrieve a tile from a GPU than from the host RAM. 
To the best of our knowledge, BLASX is the first linear algebra library that considers such 
tile cache hierarchies on multi-GPU systems.

\vspace{-0.1in}
\begin{algorithm}
\caption{The ALRU Operations}
\label{AALRU_alg}
\DontPrintSemicolon
\KwData{Tile\_Host\_Address ($HA$) and $ALRU$}
\KwResult{GPU\_Address ($GA$)}
\Fn{$Translate$ ($ALRU$, $HA$)}{
    $LRUBlock(LB) \leftarrow ALRU.HashMap(HA)$\\
    \If{$LB = \emptyset$} {
        $GA \leftarrow Malloc(TileSize)$\\
        \If{$GA = \emptyset$} {
            $GA \leftarrow ALRU.Dequeue()$\\
        }
        $ALRU.Enqueue(HA, GA)$\\
        \Return $GA$ \tcc*{new tile cache}
    } \Else { 
        \Return $LB.GA$ \tcc*{cache hit}
    }
}

\Fn{$Dequeue$ ()}{
    $LRUBlock(LBEnd) \leftarrow ALRU.end$\\
    \While{$LBEnd \neq ALRU.begin$} {
        \If{$LBEnd.Reader = 0$} {
            remove the $LBEnd$ from the $ALRU$\\
            \Return $LBEnd.GA$
        }
        \Else {
            $LBEnd \leftarrow LBEnd.Previous$
        }
    }
}

\Fn{$Enqueue$ ($HA$, $GA$)}{
    $LRUBlock(LB) \leftarrow Malloc(LRUBlock)$ \\
    $LB.HA \leftarrow HA, \mbox{ } LB.GA \leftarrow GA$ \\
    $LB.Reader \leftarrow 0$
    $ALRU.HashMapInsert(HA, GA)$\\
    $ALRU.InsertFront(LB)$
}
\end{algorithm}
\vspace{-0.15in}

To implement the L1 tile cache, we need a LRU to discard the least frequently used tiles.
Unfortunately the vanilla LRU algorithm \cite{LRU} can not accommodate the asynchronous kernel 
launches in BLASX. Consequently, we propose the Approximate Least Recent Used (ALRU) to handle asynchronicities.
Alg.\ref{AALRU_alg} presents the three basic operations in the proposed ALRU.
It adopts a reader (line 22) to track the tile usage. Specifically 
the reader of a tile is atomically incremented if a task need it. On the other hand, the 
reader is atomically decremented if a task releases the tile. We only update readers promptly 
after the synchronization point because that's the only place to inform the tile status 
(line 17 in Alg.\ref{BLASX_alg}). A reader equals 0 indicating no tasks using it; thereby 
it can be released in the ALRU. Since the runtime does not immediately synchronize readers 
in the asynchronous progress, it is possible to have nonzero reader on the least recent used 
tile. This deviates from the vanilla LRU policy by discarding the first approximate (as opposed to the exact)
least recent used tile having the zero reader, which is reflected by lines 14-18 in Alg.\ref{AALRU_alg}.
\begin{figure}[!t]
\vspace{-0.1in}
\centering
\includegraphics[width=2.0in]{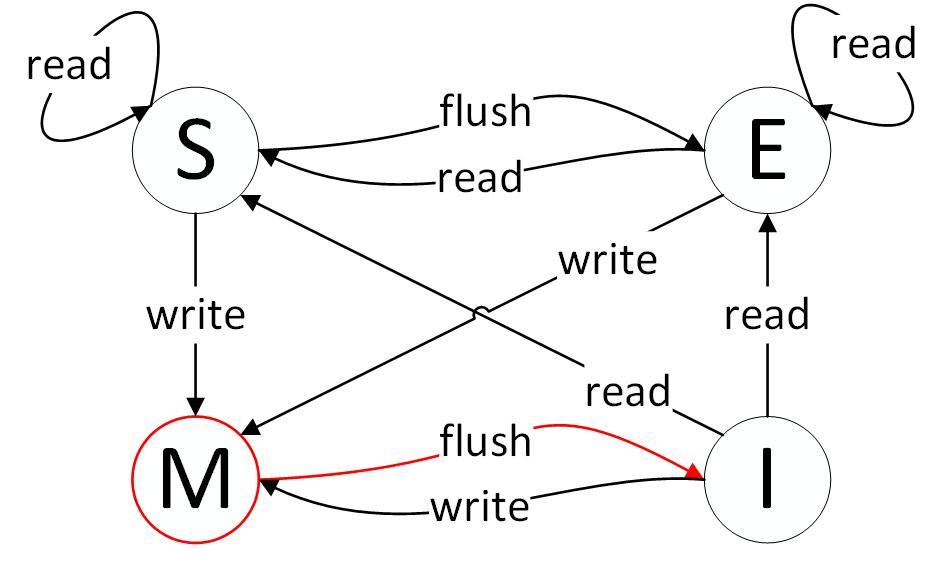}
\caption{The state transition diagram of MESI-X protocol in BLASX. The red M state is an ephemeral state, 
which immediately transits to I state by writing the tile back to host RAM.}
\label{MESI-X}
\vspace{-0.25in}
\end{figure}

To implement the L2 tile cache, we need a cache coherence protocol to ensure the consistency of shared tiles in 
multiple places. We adopt the MESI-X cache coherence protocol, a variant of MESI protocol \cite{MOESI}, to work for the BLASX's particular
context. Each ALRU associates with a specific GPU; The ALRUs all together reflect tile states in accordance with 
MESI-X protocol as follows: A tile is at E state if only a ALRU tracks it; A tile is at S state if multiple ALRUs 
track it; A tile is at I state if no ALRUs track it; A tile is at M state if a GPU writes a $\mathbf{C_{ij}}$ to it. Unlike the
regular MESI protocol, we define the M state as an ephemeral state that writes back any tile in the state to the host RAM 
and sets the tile state immediately to I. Fig.\ref{MESI-X} demonstrates the state transition diagram of the proposed MESI-X.

\vspace{-0.12in}
\subsection{Scheduling Infrastructure and Strategies}
\begin{figure}
\vspace{-0.1in}
\centering
\includegraphics[width=2.5in]{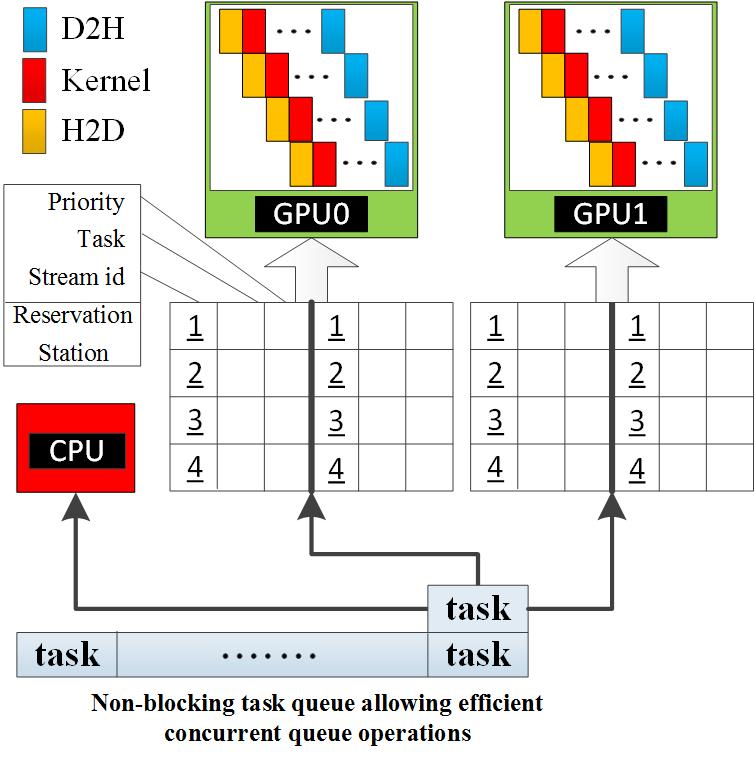}
\caption{The runtime infrastructure for our locality aware dynamic scheduler. G2H is 
the GPU to Host data transfer while H2G is the reverse.}
\label{scheduling_infrastructure}
\vspace{-0.25in}
\end{figure}

\vspace{-0.05in}
Our scheduling runtime achieves three specific goals: the proper load balancing on heterogeneous 
multi-GPUs and multi-CPUs, the efficient communication with locality aware scheduling and the
sufficient overlapping of computation and communication. Fig.\ref{scheduling_infrastructure} 
presents the scheduling infrastructure in our locality aware scheduling
runtime, which consists of 4 major components:

\vspace{-0.1in}
1) \textit{GPU Computation Thread}: a CPU thread to submit tasks for a 
specific GPU. To avoid the OS scheduling preemption, we bind the thread to a 
dedicated CPU core. The communication/computation overlapping requires at least 
2 tasks concurrently running on streams; while Wei et al.\cite{PaRSEC} demonstrate no 
performance gain when using more than 4 streams. This leads us to adopt 4 concurrent tasks to overlap
the computation/communication, which also explains the 4 streams used in Fig.\ref{scheduling_infrastructure}.

\vspace{-0.1in}
2) \textit{CPU Computation Thread}: a CPU thread to submit tasks for the rest of CPU cores. 
Peng et al. proposes the hybrid tile layout to CPU Cores and GPUs due to the inherent devices' performance 
differences \cite{hybridtile}. 
We adopt the same concept but different approach. The CPU cores dequeue one task at each time and solve 
the task with a multithreaded BLAS kernel, where the tile is further factorized.

\vspace{-0.1in}
3) \textit{Reservation Station (RS)}: a buffer designed to hold the upcoming 
tasks for a GPU. The runtime conducts work stealing
and priority scheduling on it. Each slot of a RS conveys a task priority, a task metadata, 
and a stream index.

\vspace{-0.1in}
4) \textit{Non-blocking Task Queue}: It is a non-blocking queue allowing
efficient concurrent dequeue and enqueue operations based on the algorithm proposed by 
Maged and Michael \cite{nbqueue}.

On heterogeneous systems, tasks can be executed on any computing device with load balancing 
being key to achieve optimal performance. In reality, it is possible 
to have task workload variation computed on the processors of various speeds. The two effects 
accentuate the uncertainty of processors on tasks consumption speed. One simple load balancing solution is to distribute tasks based on the inherent processors' speeds, which is the
case in PaRSEC; the actual execution time, however, changes with the GPU kernel
saturation and the actual tasks' workload. Hence this solution may lead heavy tasks, however rare to clog the slower processor(s). Our load balancing scheme leverages the task-level workload and the processors' 
real time speed to achieve the optimal performance. We treat GPUs about to entering idle states
as a sign of demand, causing the thread to dequeue tasks. The key of our dynamic 
task-scheduling runtime is that processors simultaneously pull out tasks from the 
global non-blocking task queue by their demands. Specifically, faster processors 
consume more tasks and initiate more demands while the slower processors consume fewer 
and demand less. The load is thereby adjusted according to the real time demand of individual 
processors. The perfect scenario in this case is such that processors, regardless of speed 
difference, spend identical time on task execution without idling.

\begin{figure}[t]
\centering
\includegraphics[height=1.2in]{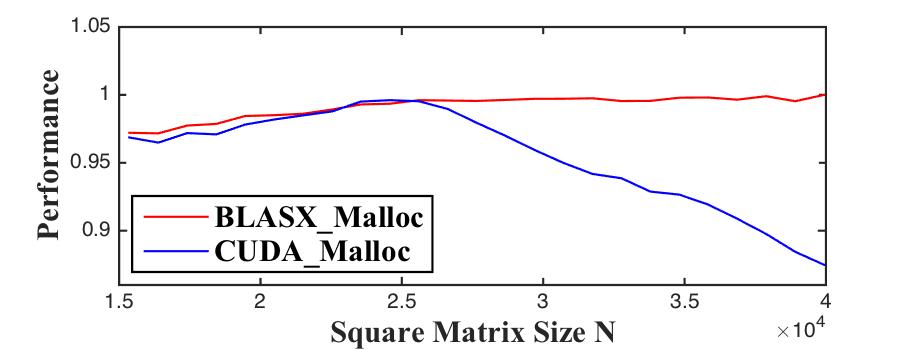}
\caption{Performance degeneration when increase the matrix size using CudaMalloc and CudaFree.}
\label{blasx_malloc}
\centering
\includegraphics[height=1.105in]{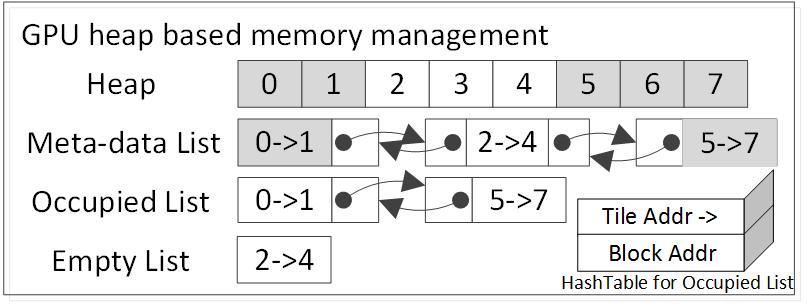}
\caption{A fast heap design to amortize the GPU memory allocation/deallocation overhead.}
\label{heap_memory}
\vspace{-0.23in}
\end{figure}

We adopt the work sharing and the work stealing to achieve this demand driven load balancing on the 
proposed runtime infrastructure. Lines 10 to 13 in Alg.\ref{BLASX_alg} indicate each GPU populates
the affiliated RS either through global task queue or work stealing. The global non-blocking task queue 
simulates the work sharing by enabling concurrent task retrieval from multi-GPUs. Since the
line 16 in Alg.\ref{BLASX_alg} is a synchronization point, a GPU will not demand tasks from
RS unless it finishes current ones. Hence GPUs that demand more attempt to pull out more tasks from
the queue. The work stealing intends to further
improve the load balancing under the situation of empty global task queue but full RS on GPUs. In 
this case, a underutilized GPU or CPU takes the initiative to steal a task from a overloaded RS,
further balancing tasks at a finer level.

Prioritizing tasks with the better temporal locality lessens unnecessary communication.
Lines 14, 15 and 19 in Alg.\ref{BLASX_alg} demonstrate task prioritization in the runtime.
At each $k$, a GPU fetches top 4 prioritized tasks from the affiliated RS. The runtime refreshes 
the priorities in RS after new tasks coming in. The runtime populates tasks' priorities
based on the extent of potential cache hits, which follows:
\vspace{-0.1in}
\begin{subequations}
 \label{tile_priority}
 \begin{align}
  priority &= \sum \limits_{k=0}^{z} \left(f(\mathbf{A_{ik}}) + f(\mathbf{B_{kj}})\right) \label{priority1} \\
  f(\mathbf{X}) &= \begin{cases} 0, \mbox{      }Otherwise\\
                                 1, \mbox{      if L2 cache hit}\\
                                 2, \mbox{      if L1 cache hit}\\
                   \end{cases}
\end{align}
\end{subequations}
The tile locality has 3 scenarios: it hits the L1 tile cache; it hits the L2
tile cache and it is located at the host RAM. In terms of communication cost, there is
no cost for L1 cache hit, and the L2 cache hit incurs less cost than retrieving the tile from the host.

\vspace{-0.12in}
\subsection{Overlapping Computation with Communication}
\vspace{-0.05in}
The CUDA stream is sequential operations executed in the issued order on the GPU with two notable properties. First, the operations on different streams can be simultaneously 
executed within the same physical device. This property enables the communication/computation 
overlapping via moving the data on one stream while executing kernels on another one. 
Secondly, streams can divide the GPU processing power between a few tasks by allocating 
segments to each task in turn. With these two properties, a tight interleaving of tasks on 
multiple streams can render the actual communication cost trivial.

L3 BLAS tile algorithms indicate that a task essentially involves $k \in [1, z]$ steps of kernel execution.
The runtime overlaps the communication/computation by interleaving tasks as follows: 
First, the RS directly maps top 4 prioritized tasks onto 4 CUDA streams. 
Second, the runtime executes each step, $k$, on all streams such that tasks in the step are computed before 
advancing to the next step (line 19-25 Alg.\ref{BLASX_alg}), which is guaranteed by the sequential execution 
feature of CUDA stream. Solving a step of tasks involves data transfer of the required tiles (if cache miss) followed 
by kernel executions. As tasks progress on multiple streams in this way, the data transfer on a stream eventually overlaps 
with the kernel execution on another one. In sum, we can consider the overall execution time to be the sum of kernel execution time, plus 
initial tile move in, and plus final tile move out. This negligible communication cost enables the input and output matrices to 
reside on the host memory. Consequently we claim our GPU operations are out-of-core.

\vspace{-0.12in}
\subsection{Amortize GPU Memory Allocation/Deallocation Overhead}
\vspace{-0.05in}
GPUs require memory allocation for tiles move-in and deallocation for tiles move-out. 
Increasing the computation scale leads to the performance deterioration due to
the significant overhead of memory allocation/deallocation. Fig. \ref{blasx_malloc} 
presents the performance degeneration with CUDA's native memory management utilities 
such as cudaMalloc and cudaFree. As a consequence, we implement a fast heap based GPU memory 
management utilities, BLASX\_Malloc, to alleviate this issue. The core concept of it is to 
amortize the allocation/deallocation overhead by adopting a big chunk of GPU memory 
as the preallocated heap.

Fig. \ref{heap_memory} presents the basic scheme of the proposed heap design,
which consists of a meta-data list, a occupied list and an empty list. A node 
in the meta-data list traces the length of memory segment and its occupation status. 
Each block of the occupied and empty list tracks the allocated segment and the free segment 
respectively. The dynamics of this heap is as follows:
During the allocation, the heap searches for the first node with enough memory in the empty list, 
which is subsequently split into two nodes. One for the occupied list recording the allocated 
memory; The other for the empty list recording the residual memory. During the deallocation, 
the runtime locates the segment in the occupied list with a hashtable. If either the 
node's left or right neighbors are contiguous with the node in terms of memory, they merge together. 
Then the segment is marked as free and placed back to empty list afterwards.
Fig. \ref{blasx_malloc} demonstrates our heap based method effectively amortizes the memory 
allocation and deallocation overhead.

\section{Performance Evaluation}
\begin{table}[!t]
\caption{The system configuration of experimental machines}
\label{Machines}
\centering
\begin{tabular}{c c c}
    \toprule
    &\multicolumn{2}{c}{\textbf{System Configuration}}\\
    \cline{2-3}
                 &      Everest            &      Makalu              \\  \midrule
    \textbf{OS}  &      CentOS v. 6.3     &      Ubuntu Server 14.04                \\
    \textbf{GPU} &      3 Kelper K40      &      2 Kelper K40 and 2 Maxwell Titan X \\
    \textbf{CPU} &      2$\times$Xeon E5 4655 V3 &      2 Xeon E5 1620 V3                  \\
    \textbf{RAM} &      64 GB DDR3        &      64 GB DDR3                         \\ 
    \textbf{CUDA}&      v 6.5             &      v 7.0                              \\
    \textbf{Compiler}   &      GCC 4.4.7         &      GCC 4.8.2                          \\
    \textbf{CPU BLAS}   &      OpenBLAS v 1.13   &      OpenBLAS v 1.13                    \\
    \bottomrule
\end{tabular}
\vspace{-0.12in}
\end{table}

In this section, we present comprehensive evaluations of our L3 BLAS routines.
We conducted the experiments on two shared memory machines Everest and Makalu, the specifications 
of which are included TABLE \ref{Machines}.

\clearpage
\begin{figure*}[t]
\centering
\includegraphics[width=0.8\textwidth]{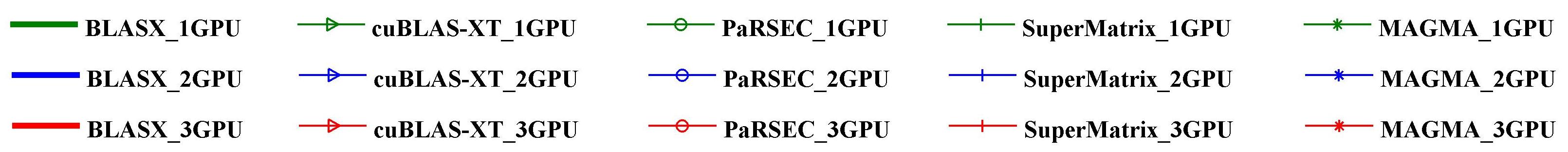}
\mbox{
\subfloat{\includegraphics[width=0.5\textwidth]{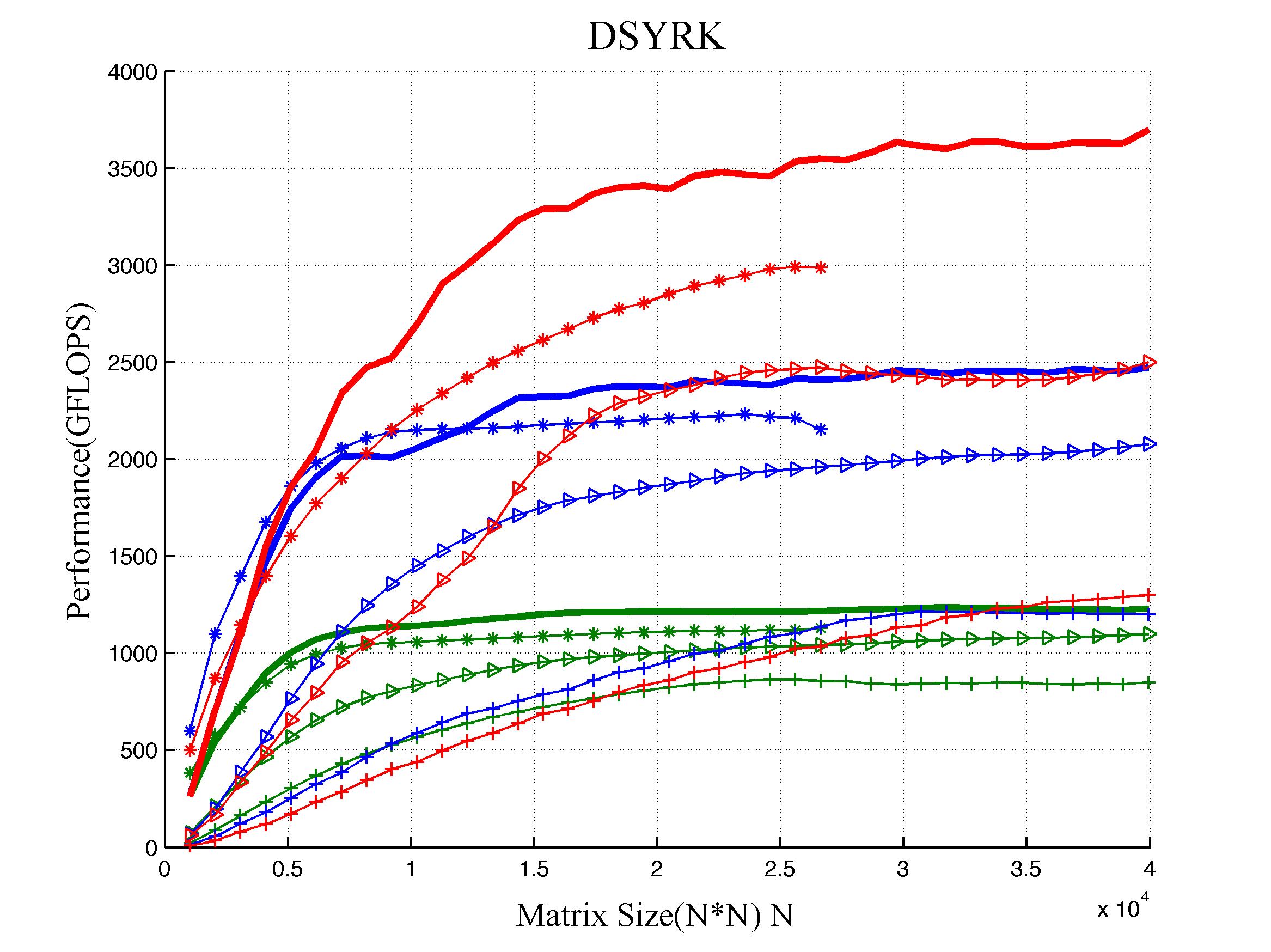}}
\subfloat{\includegraphics[width=0.5\textwidth]{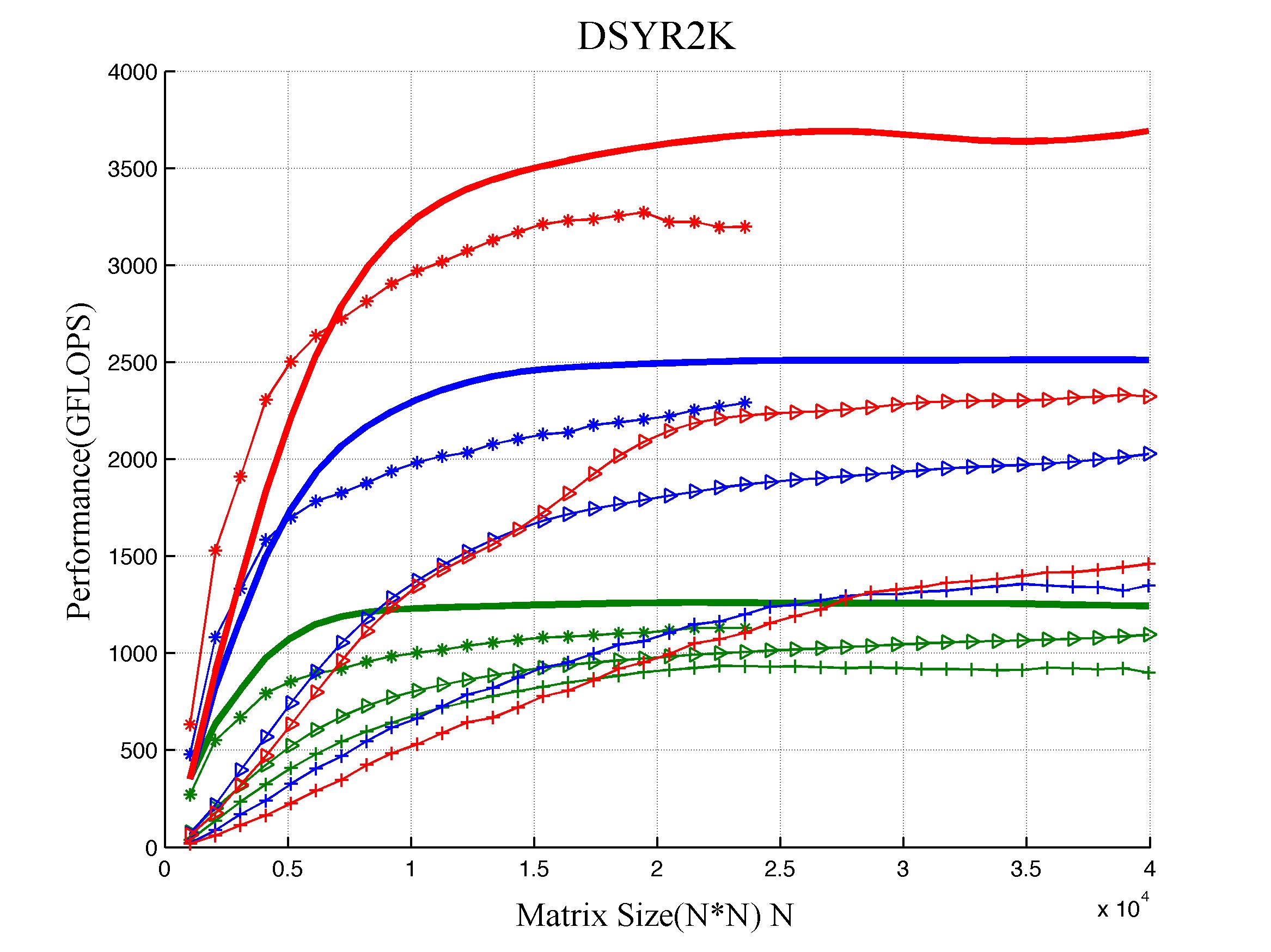}}
}
\hfil
\mbox{
\subfloat{\includegraphics[width=0.5\textwidth]{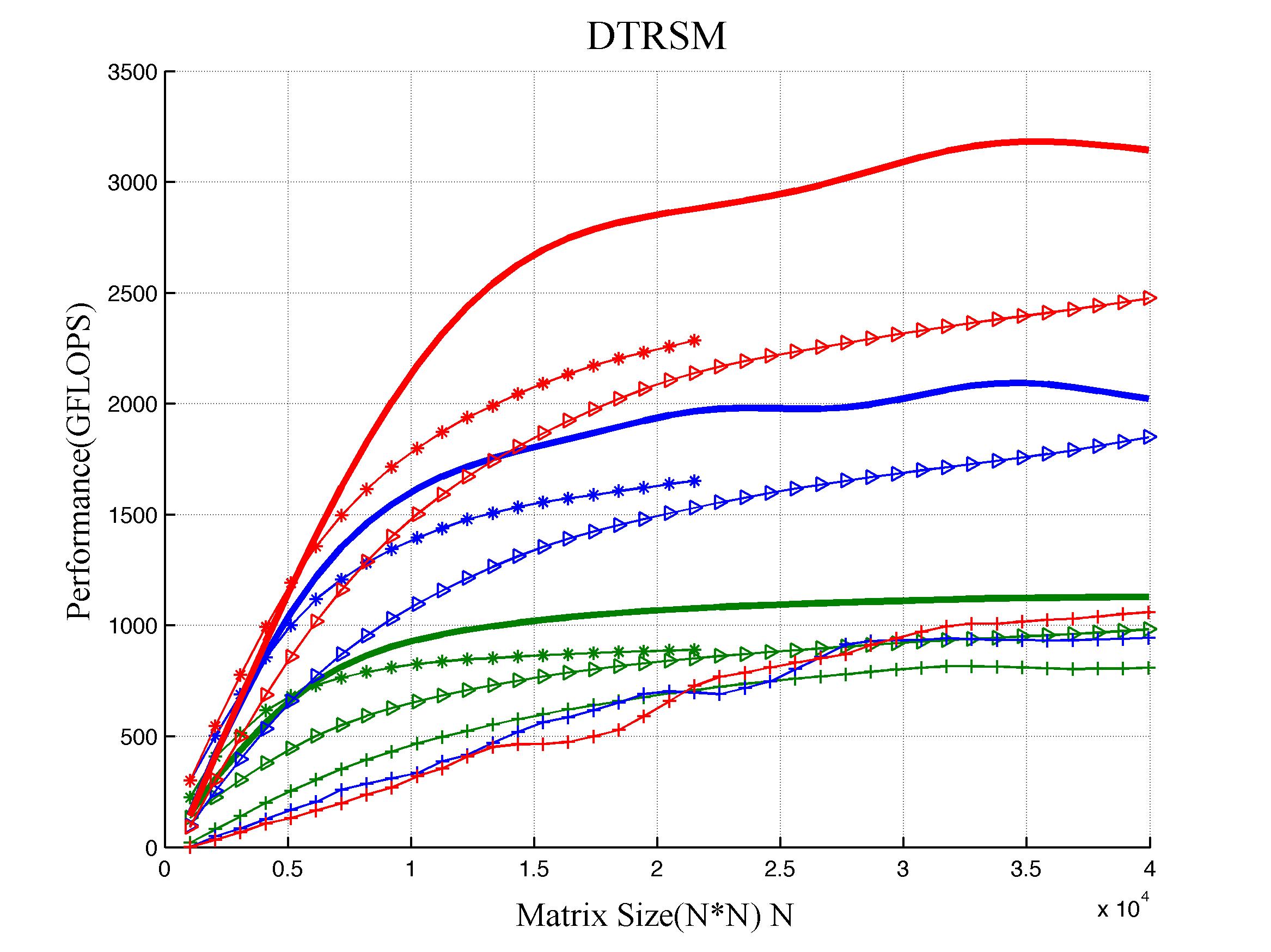}}
\subfloat{\includegraphics[width=0.5\textwidth]{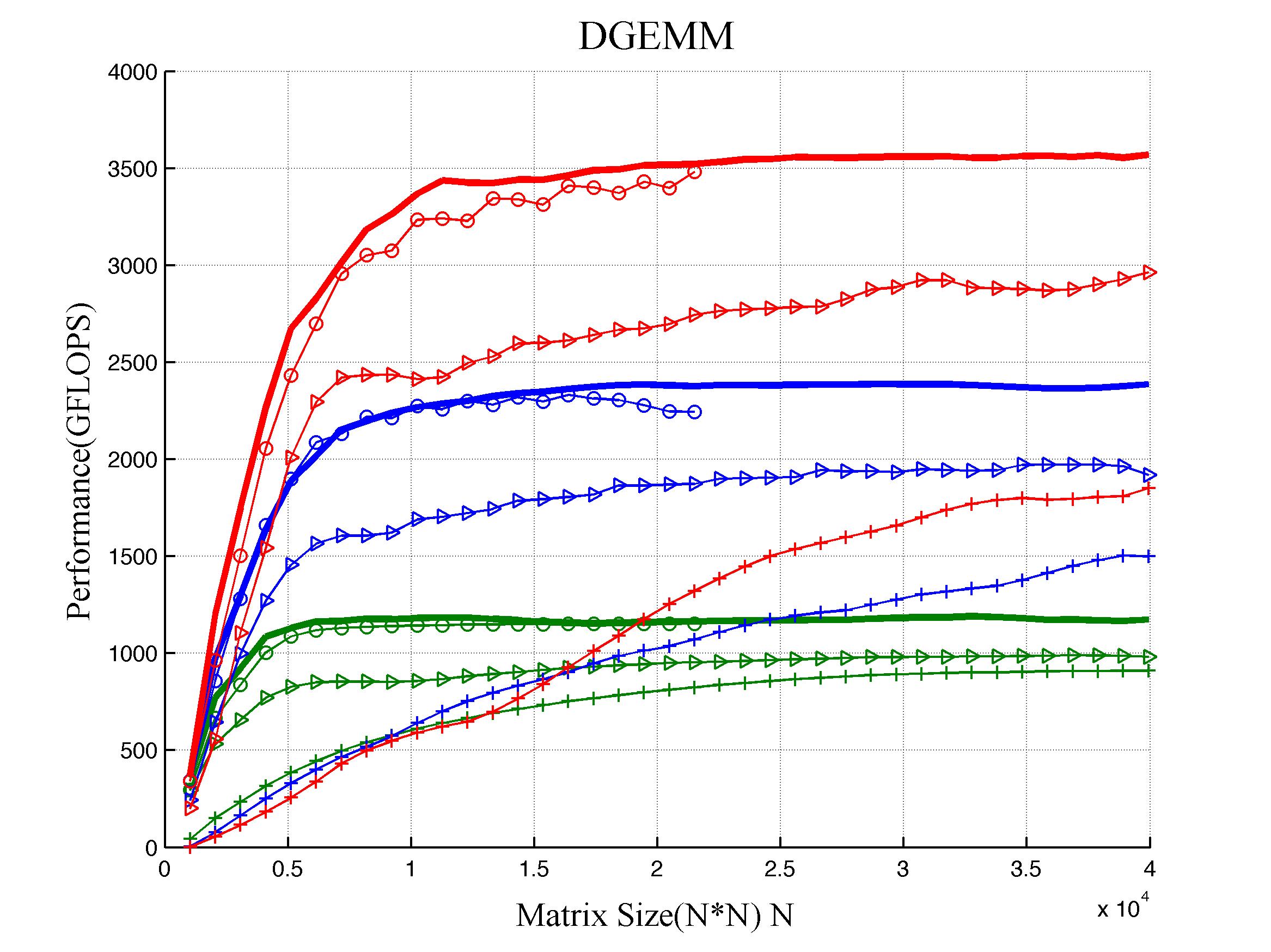}}
}
\hfil
\mbox{
\subfloat{\includegraphics[width=0.5\textwidth]{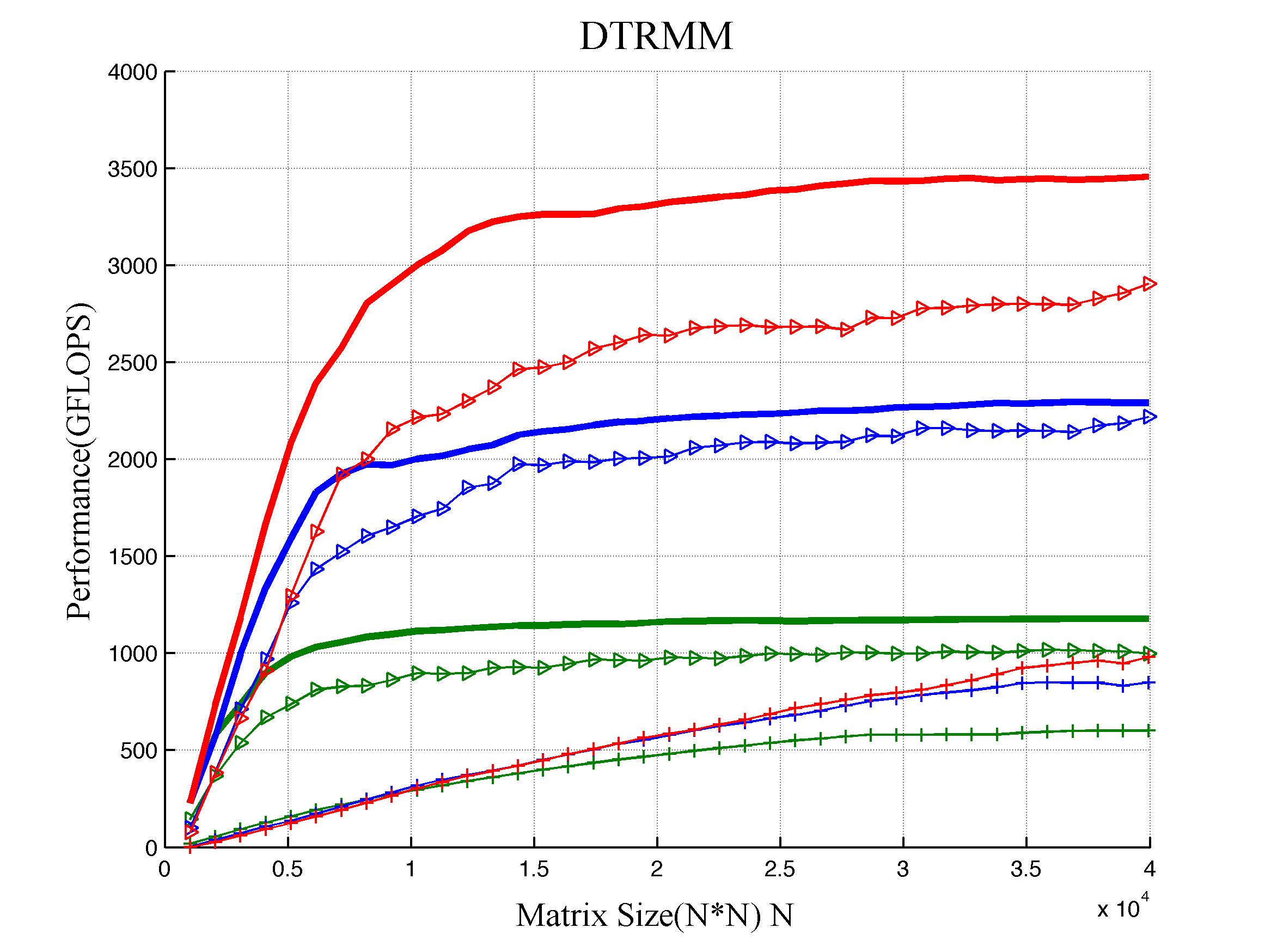}}
\subfloat{\includegraphics[width=0.5\textwidth]{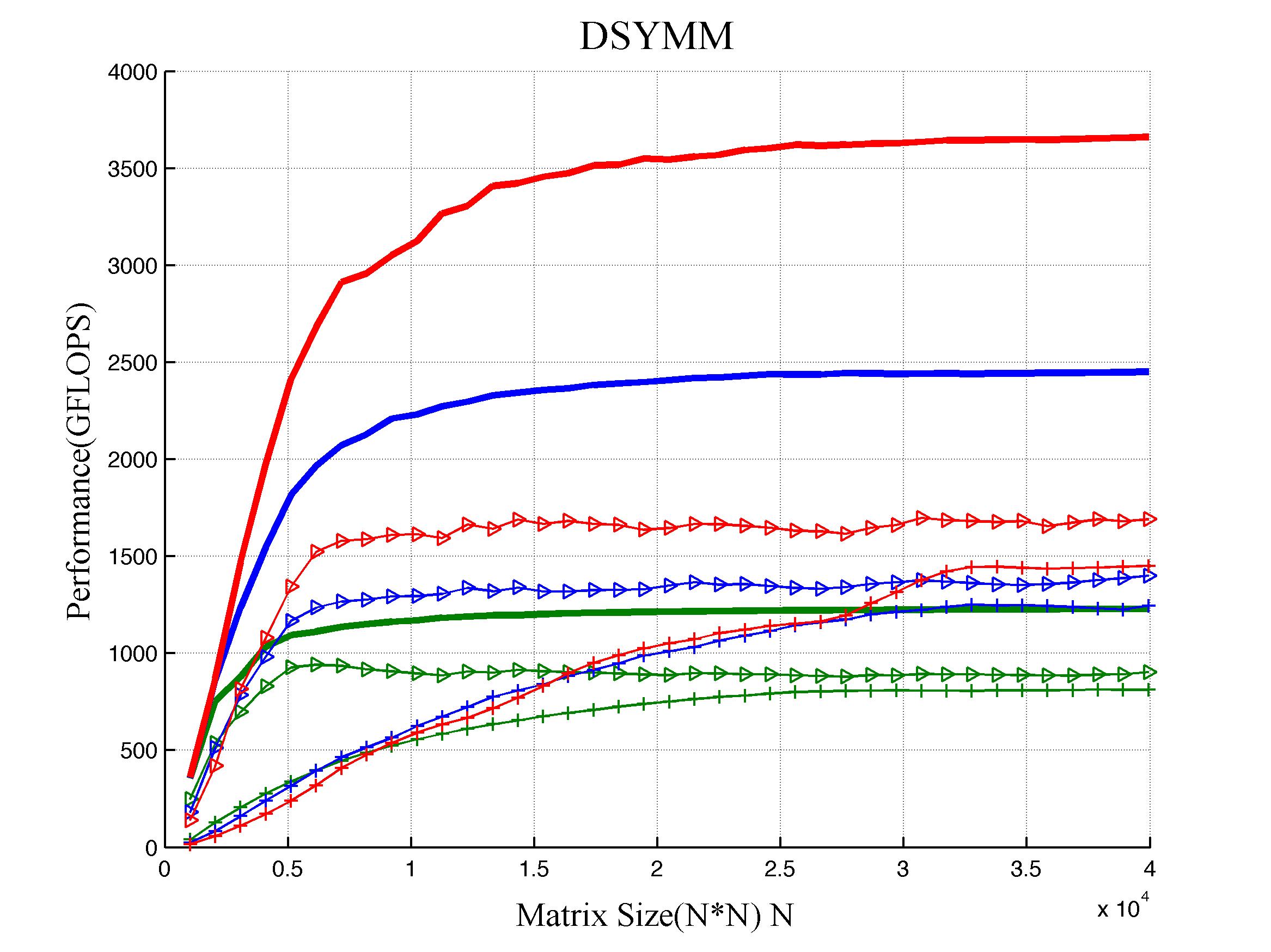}}
}
\caption{The comprehensive benchmarks of double precision L3 BLAS on Everest. BLASX
demonstrates superior performance than state-of-art commercial product cuBLAS-XT and academic 
related projects such as MAGMA and SuperMatrix. The out-of-core GPU operations in BLASX enables 
larger scale GPU computing than does the in-core GPU operations by PaRSEC and MAGMA. }
\label{benchmark}
\end{figure*}
\clearpage

\vspace{-0.12in}
\subsection{The Comprehensive L3 BLAS Benchmark}
\vspace{-0.05in}

We evaluate performance against a commercial product cuBLAS-XT and seminal 
academic libraries such as SuperMatrix, PaRSEC and MAGMA.
We setup benchmarks on the machine Everest (TABLE \ref{Machines}) using double precision L3 BLAS 
routines. The memory range of input or output matrices is page locked to expedite PCI-E transfer; 
and the time spent on page-locking is excluded from performance metric. The $alpha$ and $beta$ 
are two random float constants; other parameters such as $UPLO$, $SIDE$, $TRANS$ and $DIAG$ 
are ensured to be same in each comparison. The input matrix size $N$ starts from 1024 to 39936 at 
increments of 1024. The performance data are from the average of 3 runs; execution order of 
BLASX, cuBLAS-XT, MAGMA, SuperMatrix and PaRSEC is randomized to eliminate the potential ordering 
influence.

Fig.\ref{benchmark} demonstrates the comprehensive benchmarks on Everest. 
In single GPU benchmarks, the mean performance of BLASX converges to $92.68\%$ of 
the in-core cuBLAS DGEMM peak; whereas the average performance of PaRSEC, MAGMA, cuBLAS-XT 
and SuperMatrix attains $91.10\%$, $81.28\%$ , $79\%$ and $63.99\%$ of in-core cuBLAS DGEMM 
peak respectively. Even though PaRSEC achieves comparable performance, its GPU in-core 
operation limits PaRSEC to handle matrix sizes $N > 22528$ as the required memory, 
$22528*22528*8*3 = 12.18$ GB, is beginning to exceed the GPU RAM $12$ GB capacity. This also 
explains partial benchmarks on the MAGMA DSYR2K and DTRSM. The sufficient GPU 
communication/computation overlapping is one of predominant factors to the high performance 
of BLASX. Whereas Supermatrix follows a simple fork and join model blocking kernel launches 
until the on demand data is transferred. The non-overlapped communication in SuperMatrix 
incurs too much latency to delivery comparable performance. Hence we omit its discussion 
in the rest of paper.

\setlength{\tabcolsep}{4.5pt}
\begin{table}[!t]
\caption{The average parallel efficiency of various implementations with input square matrix $N \in [1024, 39936]$ on Everest.}
\label{parallel_efficiency}
\centering
\begin{tabular}{c c c c c c}
    \toprule
      \textbf{Routines}   & \textbf{BLASX} & \textbf{PaRSEC} & \textbf{MAGMA} & \textbf{cuBLAS-XT} & \textbf{SuperMatrix} \\  \midrule
      \textbf{DSYRK}      &    $85.54\%$   &    N/A          &     N/A        &    $64.21\%$       &      $33.33\%$           \\
      \textbf{DTRSM}      &    $81.58\%$   &    N/A          &   $77.3\%$     &    $77.27\%$       &      $30.72\%$           \\
      \textbf{DTRMM}      &    $88.99\%$   &    N/A          &     N/A        &    $82.11\%$       &      $38.96\%$           \\
      \textbf{DSYMM}      &    $90.36\%$   &    N/A          &     N/A        &    $57.96\%$       &      $43.42\%$           \\ 
      \textbf{DGEMM}      &    $93.53\%$   &  $92.85\%$      &     N/A        &    $89.85\%$       &      $46.22\%$           \\
      \textbf{DSYR2K}     &    $85.54\%$   &    N/A          &   $79.58\%$    &    $64.21\%$       &      $34.70\%$           \\
    \bottomrule
\end{tabular}
\vspace{-0.07in}
\end{table}

\begin{table}[t]
\centering
\caption{The average throughput of Direct Memory Access(DMA) engine.}
\label{DMA_throughput}
\begin{tabular}{c c c}
     \toprule
                   &           Bidirectional Host and GPU &   GPU to GPU\\  \midrule
     Throughput    &             6.54 GB/s       &    7.8 GB/s                    \\ 
    \bottomrule
\end{tabular}
\vspace{-0.15in}
\end{table}

BLASX demonstrates linear speedups and the highest scalability under multiGPU configurations.
Fig.\ref{benchmark} indicates performances increase with the matrix size and reaches a plateau 
after $N > 15000$. At the matrix size $16384$, the DSYR2K speedup of 
BLASX, cuBLAS-XT, MAGMA on two GPUs are 1.99x, 1.83x, 1.91x; and the triple GPU speedup are 
2.91, 2.16, 2.88. However, real world applications often entail small scale matrix
$N < 15000$. Measuring parallelizations at various matrix sizes is more convincing than
concluding the speedup at a particular matrix size. Since the parallel efficiency is a 
performance metric to describe the scalability at a specific problem size $N$,
we calculate the average parallel efficiency based on 39 matrix sizes $N \in [1024, 39936]$ 
to yield a global insight about scalabilities at various matrix sizes; and we adopt forward
padding to partial benchmarks in MAGMA and PaRSEC. In TABLE \ref{parallel_efficiency}, 
BLASX outperforms the second 
best alternatives at the average rate of $5\%$. Particularly BLASX DSYMM is $32.4\%$ 
higher than the second best implementation by cuBLAS-XT. There are 4 major factors that contribute
to the success of BLASX, which are 1) the demand-driven load balancing, 2) the seamless 
GPU occupancy, 3) the significantly less communication volume and 4) the efficient GPU-GPU 
P2P communication. We investigate each factors as follows.

\begin{figure}[!t]
\centering
\subfloat[DGEMM]{\includegraphics[height=.55in]{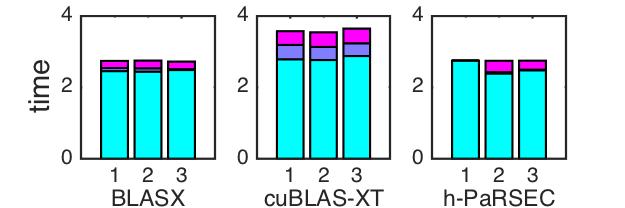}
\label{gemm_cc}}
\hfil
\subfloat[DSYMM]{\includegraphics[height=.55in]{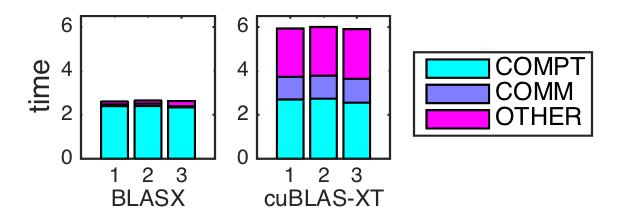}
\label{symm_cc}}
\\
\subfloat[DTRSM]{\includegraphics[height=.55in]{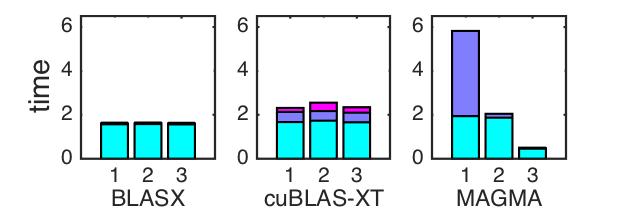}
\label{trsm_cc}}
\hfil
\subfloat[DTRMM]{\includegraphics[height=.55in]{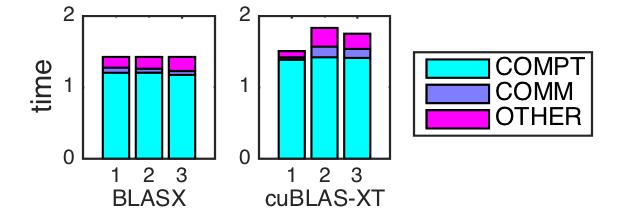}
\label{trmm_cc}}
\\
\subfloat[DSYR2K]{\includegraphics[height=.55in]{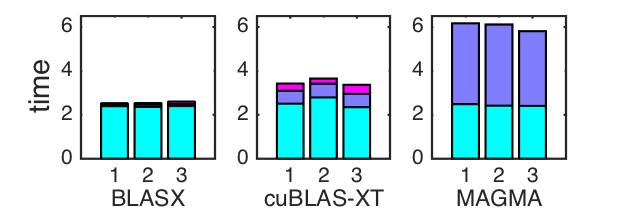}
\label{syr2k_cc}}
\hfil
\subfloat[DSYRK]{\includegraphics[height=.55in]{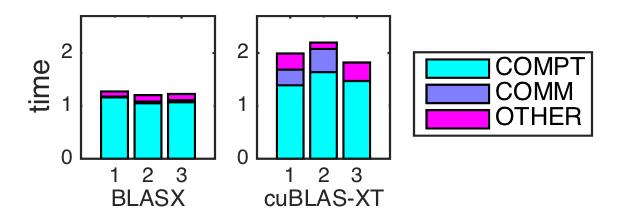}
\label{syrk_cc}}
\caption{The execution time profile (classified into COMPT, COMM and OTHER) at the square matrix size $N = 16384$ on Everest. 
The horizontal indices 1, 2, 3 represents GPU\_1, GPU\_2 and GPU\_3.}
\label{time_profile}
\vspace{-0.3in}
\end{figure}

First, our demand-driven dynamic load balancing is key to the BLASX high performance.
In Fig.\ref{time_profile}, we dissect each GPU execution profiles into 3 major
components---the computation (COMPT), the unoverlapped communication (COMM), the 
synchronization and free gaps among kernel launches (OTHER). A typical ideal scheduler 
allows each GPUs, regardless of differences, to spend identical time during the execution. 
Fig.\ref{time_profile} indicates that dynamic schedulers employed by BLASX and PaRSEC 
are better than static schedulers by MAGMA and cuBLAS-XT. For example, the average elapsed time
differences between the fastest GPU and the slowest GPU of cuBLAS-XT and BLASX are $0.2961$
and $0.0391$ seconds; and the same metric for MAGMA (only count COMM) and BLASX is $0.7837$ 
and $0.0457$ seconds. The DGEMM of PaRSEC and BLASX attains comparable performance of $0.0252$ 
and $0.0285$ seconds.

Second, the seamless GPU occupancy enables BLASX to fully saturate each GPU. In Fig.\ref{time_profile},
BLASX demonstrates the least none-computation cost (OTHER+COMM). The communication is counted 
toward latency if it is not overlapped. The average communication latency 
of BLASX is 0.0575s while cuBLAS-XT is 0.4917s. The difference is largely due to the significantly reduced communication volume and the seamless stream-level overlapping and kernel launches. 
OTHER includes synchronization latency and the minor GPU idle gaps among kernel launches. The synchronization 
is necessary to ensure the mathematical rigorousness; and idle gaps depend on the tightness 
of kernel launches on multistreams. Increasing streams, as demonstrated by Wei et al \cite{PaRSEC}, 
improves the GPU saturation by reducing those gaps. BLASX dynamically interleaves tasks over 
multiple streams, while cuBLAS-XT adopts two.

Third, the hierarchical tile caches in BLASX dramatically diminish the communication volume.
According to TABLE \ref{communication_volume}, the average communication volume of cuBLAS-XT, 15143 MB, 
is 2.95 times of BLASX, 5132 MB. The L1 tile cache of BLASX exploits the tile temporal locality 
to minimize global communication, which is not the case for cuBLAS-XT. The increasing 
gaps among the GPU clock frequency (1.43 TFLOPS on K40c), GPU memory bandwidth (288 GB/sec on K40c)
and PCI-E bandwidth (31.51 GB/s v4.0 x16) identify GPU off-chip memory access extremely 
expensive. Consequently, the excessive data transfer of cuBLAS-XT incurs a huge latency penalty to its 
performance. The DGEMM data also indicates that BLASX (6219 MB) saves 12\% communication
over PaRSEC (6961 MB).

\setlength{\tabcolsep}{4.5pt}
\begin{table}[t]
\centering
\caption{The communication volume(in MB) of L3 BLAS routines at the input square matrix size $N = 16384$.}
\label{communication_volume}
\begin{tabular}{c c c c c c c}
    \toprule
    \setlength{\tabcolsep}{1em}
    & \multicolumn{3}{c}{\textbf{DGEMM}} & & \multicolumn{2}{c}{\textbf{DSYMM}} \\
    \cline{2-4}  \cline{6-7}
    MB             &      BLASX                 &  cuBLAS-XT & h-PaRSEC  & &   BLASX                          &  cuBLAS-XT  \\  \midrule
    \textbf{GPU1}  &       6895                 &    24528   &      7563 & &   5628                           &    18865         \\
    \textbf{GPU2}  & \textcolor{red}{1811}+4768 &    24243   &      7160 & &   \textcolor{red}{1249}+5318     &    18865         \\
    \textbf{GPU3}  & \textcolor{red}{721}+4462  &    24243   &      6160 & &   \textcolor{red}{343}+3758      &    18102         \\
    \bottomrule
    & \multicolumn{3}{c}{\textbf{DTRSM}} & & \multicolumn{2}{c}{\textbf{DTRMM}} \\
    \cline{2-4}  \cline{6-7}
    MB             &      BLASX                  &  cuBLAS-XT &    MAGMA  & & BLASX                           &  cuBLAS-XT  \\  \midrule
    \textbf{GPU1}  &     2751                    &    12985   &    15130  & &   4966                          &  8885       \\
    \textbf{GPU2}  & \textcolor{red}{981}+4622   &    12792   &    2768   & &   \textcolor{red}{1694}+4353    &  2020       \\
    \textbf{GPU3}  & \textcolor{red}{300}+2575   &    12876   &    1587   & &   \textcolor{red}{327}+2365     &  8922       \\
    \bottomrule
    & \multicolumn{3}{c}{\textbf{DSYR2K}} & & \multicolumn{2}{c}{\textbf{DSYRK}} \\
    \cline{2-4}  \cline{6-7}
    MB             &      BLASX                 &  cuBLAS-XT &    MAGMA & &  BLASX                       &  cuBLAS-XT  \\  \midrule
    \textbf{GPU1}  &       7281                 &    15910   &    5738  & &   4278                       &  12314      \\
    \textbf{GPU2}  & \textcolor{red}{2231}+5905 &    15910   &    5720  & &   \textcolor{red}{1363}+3410 &  12834      \\
    \textbf{GPU3}  & \textcolor{red}{1308}+2969 &    15900   &    5720  & &   \textcolor{red}{1213}+2536 &  11492      \\
    \bottomrule
\end{tabular}
\begin{tablenotes}
            \item[a] \mbox{\textdagger }  The red represents the volume of GPU to GPU communication.
            \item[b] \mbox{\textdaggerdbl } The black represents the volume of bidirectional Host to Device communication.
            \item[b] \mbox{\textbullet } The Peer access is only available between GPU2 and GPU3 on the machine Everest.
\end{tablenotes}
\vspace{-0.1in}
\end{table}

Finally, reducing the CPU-GPU communication to the GPU-GPU communication further improves 
the communication efficiency of BLASX. One of the defining features of BLASX is the implementation of L2 tile 
caches, the purpose of which is to retrieve tiles from hardware adjacent GPUs under L1 misses. 
TABLE \ref{DMA_throughput} justifies our L2 tile cache 
proposal indicating that the average GPU-GPU data transfer is 19.27\% faster than the CPU-GPU 
transfer. We highlight the GPU-GPU communication volume in TABLE \ref{communication_volume}.
The interGPU communication only happens between GPU2 and GPU3 as only them share the same PCI-E switch
on Everest. We expect the efficient GPU-GPU communication eventually dominating 
with more GPUs available on the system.

\begin{figure}[t]
\centering
\includegraphics[height=1.2in]{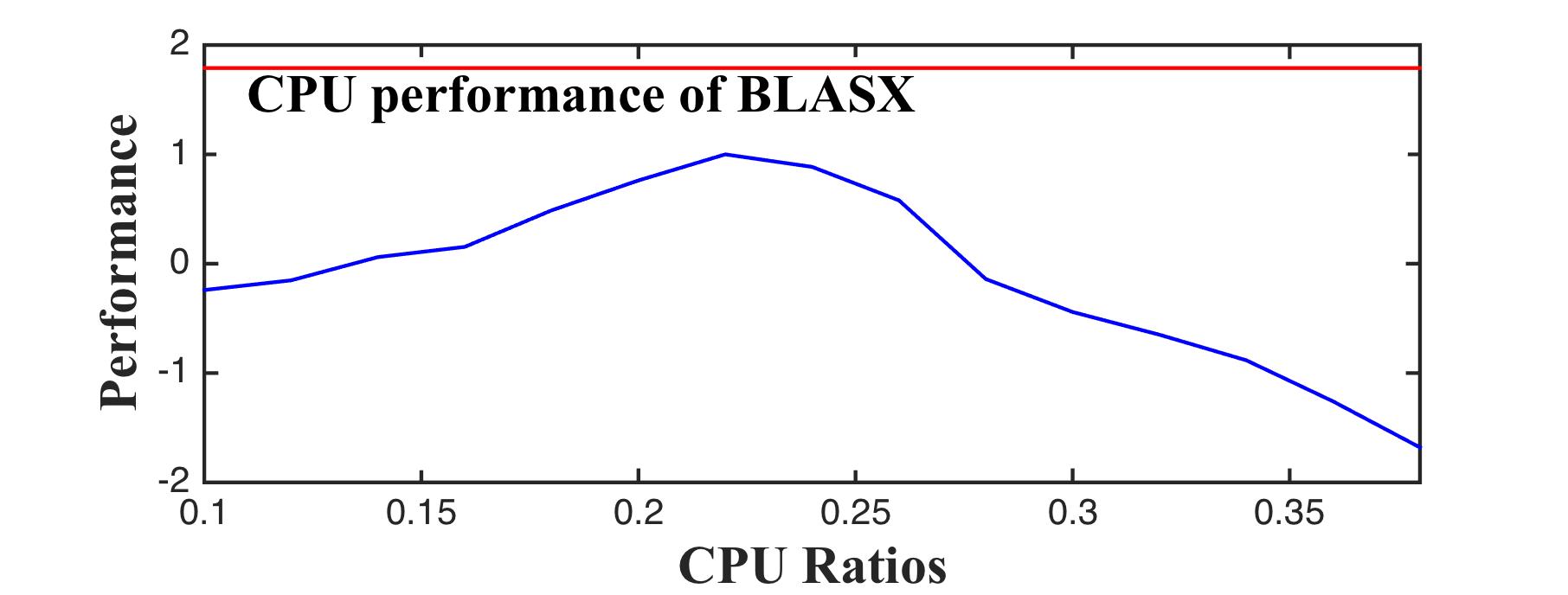}
\caption{The CPU performance of cuBLAS-XT and BLASX at various CPU ratios.}
\label{cpu_ratio}
\centering
\includegraphics[height=1.2in]{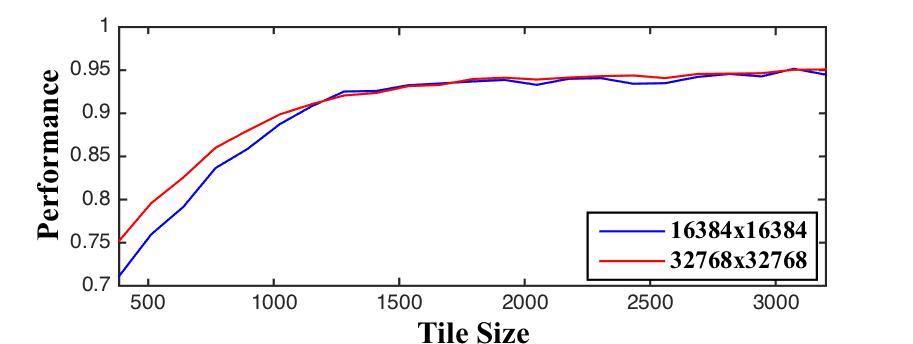}
\caption{The performance variations with respect to different tile sizes on Everest.}
\label{tile_size_tuning}
\vspace{-0.3in}
\end{figure}

Fig.\ref{cpu_ratio} presents the DGEMM CPU performance of cuBLAS-XT and BLASX on the machine Makalu.
We sample the CPU contribution by taking the difference of CPU enabled DGEMM to CPU disabled DGEMM 
under same scenarios. Our runtime automatically assign tasks to CPU; hence the performance is represented 
with a horizontal line. Fig.\ref{cpu_ratio} indicates the CPU contribution of BLASX is 78\% faster than 
the best case in cuBLAS-XT. The downtrend in the figure also suggests an improperly chosen CPU ratio 
overloads the CPU at the expense of GPUs.

\vspace{-0.12in}
\subsection{The Only Tuning Parameter---Tile Size}
\vspace{-0.05in}

We strive to develop a portable L3 BLAS library that delivers the optimal performance on different platforms 
with the least effort from users. In BLASX, the tile size is the only tuning parameter. Generally a large tile 
size reduces the degree of parallelism while a small tile size poorly saturates the GPU and PCI-E. The ideal 
tile size is a result of trade-off among GPU saturation, PCI-E efficiency and available parallelism. 
Fig. \ref{tile_size_tuning} demonstrates the impact of tile size on the overall DGEMM performance at 
two matrix sizes. The performance increases with tile size and reaches a plateau eventually. In this case, 
we choose the tile size $1024\times1024$ for benchmarks on Everest.

\vspace{-0.12in}
\subsection{Applications}
\vspace{-0.05in}

BLASX intends to provide multiGPU performance boost by directly replacing the CPU based 
alternatives. Libraries such as MAGMA, h-PaRSEC and StarPU requires extensive changes on 
the legacy code to comply with standards while cuBLAS-XT requires commercial licenses without
delivering a proportional increase of performance. In this section, we present a few applications of BLASX as follows:

\begin{inparaenum} [\itshape a\upshape)]

\item Caffe \cite{caffe} is one of the most popular deep learning frameworks nowadays, in which 
BLAS computes the image convolution \cite{img_conv}, forwards and backwards passes of densely connected
layers \cite{ANN}. We built a CPU version of Caffe and changed the BLAS linkage to BLASX.
An Artificial Neural Network (ANN) of architecture, 3072$\rightarrow$16384$\rightarrow$16384$\rightarrow$10, 
was trained with CIFAR-10 dataset \cite{cifar10} to classfiy images in 10 categories. We used
Caffe's benchmark utilities to measure the average elapsed time for a forward and backward passes
on the machine Makalu. The experiment data demonstrates that BLASX accelerates ANN 
training up to 2.48 W.R.T Caffe GPU training and 62.3 W.R.T Caffe CPU training. In terms
of model parameters, the Caffe's single GPU training can accommodate up to $1.5*10^{9}$ parameters
on a 12 GB GPU. BLASX, however, enables a model of $3.2*10^{10}$ parameters on a multiGPU server
with 256 GB host RAM. Please refer to our poster \cite{ANN_BLASX} for more details\footnote{The CPU
activation of Caffe is a single thread function, which is not efficient. We modified the function to be 
multithreaded to benchmark the results.}.

\item MATLAB, R and Octave are renowned technical computing languages widely used in academia
and industry. These scientific languages offload the primitive matrix or vector operations to the BLAS to
ensure the performance. Users can integrate BLASX with MATLAB by simply exporting the 
environment variable BLAS\_VERSION to the location of BLASX.
Although MATLAB has the GPU computing toolbox, it requires users to manually manage the data on the GPU
yet without multiGPU support. BLASX elegantly resolves these issues with 
its underlying dynamic runtime. Table \ref{MATLAB} presents a few exemplary instances of MATLAB's 
SIMULINK libraries while using BLASX. There are additional vast applications of BLASX such as finding the shortest path in a graph, 
the topology optimization\cite{topology_optimization} and finite element analysis in structure mechanics \cite{FEM}.

\begin{table}[t]
\centering
\caption{The exemplary performance boost of MATLAB SIMULINK libraries while using BLASX on Everest.}
\label{MATLAB}
\begin{tabular}{c c c}
     \toprule
      command     &  Description                &   Speedup                         \\  \midrule
      $\mathbf{A}*\mathbf{B}$ & matrix multiplication in single precision. & 12.75x \\
      $\mathbf{A}*\mathbf{B}$ & matrix multiplication in double precision. & 8.27x  \\
      nnmf     & factorize the $\mathbf{A}$ into nonnegative factors $\mathbf{W}$ and $\mathbf{H}$. & 6.72x       \\ 
      rotatefactors  & rotates the $\mathbf{A}$ to maximize the varimax criterion.      & 5.83x                   \\
      lsqlin   & solves the linear system in the least-squares sense & 3.09x \\
    \bottomrule
\end{tabular}
\vspace{-0.2in}
\end{table}

\end{inparaenum}

\section{Conclusion}
Existing L3 BLAS libraries such as PaRSEC, MAGMA, Supermatrix, cuBLAS-XT suffers from issues such as backward compatibility,
insufficient communication/computation overlapping, inefficient communication and poor scalability. In this paper, 
we design and implement BLASX, a suite of L3 BLAS, that delivers the best L3 BLAS performance on heterogeneous multiGPU systems. 
We introduce a novel two level hierarchical tile cache to reduce the global communication; Our locality aware scheduling 
runtime perfectly balances the load across heterogeneous GPUs and CPUs. We optimized communication/computation 
overlapping on streams to renders trivial communication cost; thereby BLASX computes in a out-of-core fashion that insures input and output 
data always remains on the host RAM. Extensive benchmarks demonstrate that BLASX consistently outperforms the leading industrial and academia related projects
in terms of performance, scalability, and communication efficiency. More importantly, BLASX requires the least effort from users to integrate with
the vast amount of legacy BLAS based applications.


\end{document}